Quantum Analogical Modeling with Homogeneous Pointers

Royal Skousen <royal_skousen@byu.edu>
Department of Linguistics, Brigham Young University, Provo, Utah 84602 USA

6 May 2010

*Abstract*

Quantum Analogical Modeling (QAM) works under the assumption that the correct exemplar-based description for a system of behavior minimizes the overall uncertainty of the system. The measure used in QAM differs from the traditional logarithmic measure of uncertainty; instead QAM uses a quadratic measure of disagreement between pairs of exemplars. (This quadratic measure parallels the squaring function holding between the amplitude and the probability for a state function in quantum mechanics.) QAM eliminates all supracontexts (contextual groupings of exemplars) that fail to minimize the number of disagreements. The resulting system thus distinguishes between homogeneous and heterogeneous supracontexts and uses only exemplars in homogeneous supracontexts to predict behavior. This paper revises earlier work on QAM (in 2005) by showing that homogeneity for a supracontext can be most simply determined by discovering whether there are any heterogeneous pointers between any of the supracontext's exemplars. A pointer for a pair of exemplars is heterogeneous whenever those two exemplars are found in different subcontexts of the supracontext and take different outcomes.

DETERMINING HOMOGENEITY FOR SUPRACONTEXTS

The linguistically motivated theory of Analogical Modeling (AM) proposes that the probabilistic nature of language behavior can be accurately modeled in terms of the simultaneous analysis of all possible generalized contexts defined by a given context for which we are interested in predicting the behavior. These generalized contexts are called supracontexts. The important restriction in AM is that only supracontexts that are homogeneous in behavior can be used to predict behavior; this restriction is equivalent to saying that one must use supracontexts that permit no increase in uncertainty. In the quantum mechanical version of AM, called Quantum Analogical Modeling (QAM), the amplitude for each homogeneous supracontext is proportional to its frequency of occurrence, with the result that the probability of selecting one particular supracontext to predict the behavior of the given context is proportional to the squared frequency of that supracontext.

The fundamental question in AM (and in QAM) is how to determine the homogeneity for supracontexts. There have been three basic approaches, and for each one a separate computational procedure has been devised. Initially, in *Analogical Modeling of Language* (Skousen 1989), I defined a system of directional pointers between exemplars in the data set, identifying for each ordered pair of exemplars whether there was a change in outcome (a disagreement) or the same outcome (an agreement). For a given supracontext, I determined the number of disagreements for that supracontext and then compared that number with the total number of disagreements for all the subcontexts that made up that supracontext. In section 2.2 of Skousen 1989, I showed how homogeneity could be defined in terms of minimizing the number of disagreements – namely, by choosing only those supracontexts for which there was no increase in disagreement when comparing the supracontext against its subcontexts. If the number of disagreements was the same, then the supracontext was homogeneous; if there were more disagreements in the supracontext, then the supracontext was heterogeneous. The resulting computer program was computationally intensive and resulted in an exponential running time and memory. Even so, this program had the conceptual advantage of relying on directional pointers, thus using a simple quadratic measure of uncertainty to determine homogeneity.

A second approach to determining homogeneity has been to find other supracontextual properties equivalent to minimizing uncertainty that would be more simple to state and would not involve actual calculation of the number of disagreements. For instance, one equivalent characterization, described in section 13.2 of *Analogy and Structure* (Skousen 1992), stated that a deterministic supracontext is automatically homogeneous, while a nondeterministic supracontext is homogeneous only if all the exemplars are found in a single subcontext of the supracontext. Later, in my first <arXiv.org> paper, "Analogical Modeling and Quantum Computing" (Skousen 2000, published as Skousen 2002), I came up with a single property for determining the heterogeneity of a given supracontext, namely, a supracontext is heterogeneous if and only if there is a plurality of subcontexts and a plurality of outcomes for the exemplars in that supracontext. Most importantly, this property could be determined for a given supracontext independently of determining the heterogeneity for any other supracontext. Such a global property permitted one to use the simultaneity of quantum computing to determine heterogeneity for an exponential number of supracontexts in linear time and memory (but with the restriction that the qubits take only 0 and 1 as their states). This approach was precisely described using quantum computing in my second <arXiv.org> paper, "Quantum Analogical Modeling: A General Quantum Computing Algorithm for Predicting Language Behavior" (Skousen 2005). An abbreviated version of that paper appeared in Skousen 2007.

In this paper, my third on the quantum computing of Analogical Modeling, I return to my original measure of disagreement and use it to determine whether a directional pointer between exemplars in the same supracontext is homogeneous or heterogeneous. Basically, if a pointer from one exemplar to another leads to a change in outcome and a change in subcontext (that is, if the pointer crosses a subcontextual boundary within the supracontext and at the same time leads to a different outcome), then that pointer is heterogeneous. A supracontext will be considered homogeneous if there are no heterogeneous pointers in that supracontext. The occurrence of any heterogeneous pointers in a supracontext will reduce the amplitude of that supracontext to zero, thus making the supracontext's exemplars inaccessible when observation of the system occurs. This means that none of the directional pointers between the exemplars in a heterogeneous supracontext will be accessible to observation. On the other hand, the amplitude of a homogeneous supracontext will be equal to the number of exemplars in the supracontext and the probability of selecting a homogeneous supracontext will be determined by the number of directional pointers between pairs of exemplars. Every exemplar in a homogeneous supracontext will be connected to every other exemplar (and to itself) by a directional homogeneous pointer; and when observation occurs, one of those homogeneous pointers will be randomly selected and it will point to the predicted outcome.

It turns out that the random selection of a homogeneous pointer can be done in one step, unlike the two-step procedure implicit in quantum mechanics; that is, we randomly select any one of the accessible pointers that occur in any of the homogeneous supracontexts. We don't have to first choose one of the homogeneous supracontexts (according to its exemplar frequency squared) and then randomly choose one of the exemplars in that supracontext.

MEASURING UNCERTAINTY IN ANALOGICAL MODELING

The normal approach for measuring the uncertainty of rule systems is Shannon's information, the logarithmic measure more commonly known as the entropy $H$, defined as $H = -\sum p_j \log_2 p_j$, where $p_j$ is the probability of an outcome $\omega_j$ occurring. This measure can be given a natural interpretation (as the average number of yes-no questions needed to determine the correct outcome of a given contextual specification). Unfortunately, $H$ has some disadvantages: (1) entropy is based on the notion that one gets an unlimited number of chances to discover the correct outcome; (2) the entropy for continuous probabilistic distributions is infinite; even the entropy density is infinite for continuous distributions, and an unmotivated definition for entropy density must be devised, one that sometimes gives negative measures of entropy density (see the discussion in section 3.8 of Skousen 1992).

A more plausible and simpler method for measuring uncertainty is a quadratic one, the disagreement $Q$, defined as $Q = 1 - \sum p_j^2$ (again, $p_j$ is the probability of an outcome $\omega_j$ occurring). This measure has a natural interpretation: It represents the probability that two randomly chosen instances of a context disagree in outcome. The disagreement is based on the much more plausible restriction that one gets a single chance to guess the correct outcome rather than an unlimited number of guesses.

Corresponding to $Q$, we can define the agreement $Z$, where $Z = \sum p_j^2$. Interestingly, the agreement density $Z'$ exists (and is positive finite) for virtually all continuous probabilistic distributions. For instance, given a univariate continuous distribution $f(x)$, then $Z' = \int f^2(x)\,dx$. In fact, the agreement density can be used to measure the certainty of continuous distributions for which the variance (the traditional measure of dispersion) is undefined (see sections 3.1–3.4 of Skousen 1992). But most importantly, this measure of certainty, $Z$, and certainty density, $Z'$, shows an important connection to quantum mechanics, namely, the squaring function holding between the amplitude and the probability for a state function, which gives the probability of occurrence for a given state in the state function. Here the certainty $Z$ measures the probability of agreement between random occurrences of the states as defined by the state function.

PREDICTING THE OUTCOME USING HOMOGENEOUS POINTERS

The following discussion depends considerably on Skousen 2005 (the <arXiv.org> paper "Quantum Analogical Modeling: A General Quantum Computing Algorithm for Predicting Language Behavior") and assumes familiarity with the linguistic example described in section 1.2 of that paper (and used throughout that paper to show how QAM works). In addition, sections 2.1–2.5 of that paper show how to set up the qubits necessary for doing QAM. Most importantly, fundamental operators are also defined and from them more specific operators are derived, all of which will be assumed in this paper. In particular, in section 2.4 of that paper, the difference vector $\mathbf{D}^{[j]}$ of length $m$ is set up for each data item $j$ in the dataset (each data item is an exemplar); this difference vector compares the data item $j$ with the given context for which we are trying to predict the behavior, thus showing which variables agree and which do not.

One major difference in this paper, when compared with Skousen 2005, is that we use square arrays of qubits rather than vectors of qubits in deriving the analogical set. Instead of dealing with individual exemplars, we now consider pairs of exemplars and determine whether the directional pointers between those pairs of exemplars are homogeneous or heterogeneous. This ends up making the quantum procedure quadratic in time and memory, but still polynomial and thus tractable.

To begin with, then, let us analyze each data exemplar in terms of its contextual difference with the given context, *oma*. Following section 2.4 of Skousen 2005, we have the contextual difference vector **D**:

| *oms* | *gfa* | *cms* | *cma* | *omn* | *gfr* |
|---|---|---|---|---|---|
| *oma* | *oma* | *oma* | *oma* | *oma* | *oma* |
| 001 | 110 | 101 | 100 | 001 | 111 |

But we now convert this to a pairwise comparison of the individual contextual differences, resulting in the contextual difference array **V**×**V** (or more simply **V**$^2$); in other words, we determine whether the exemplars are in the same subcontext or not. This array will be the same for each supracontext:

from $j = 1$ to $m$ do
    from $j' = 1$ to $m$ do
        IDENTITY ($D^{[j]}$, $D^{[j']}$, $X^2_{jj'} = 1$, $V^2_{jj'} = 1$)

**V**$^2$ (1 means the subcontexts are different, 0 the same)

|  | *001* | *110* | *101* | *100* | *001* | *111* |
|---|---|---|---|---|---|---|
| *001* | 0 | 1 | 1 | 1 | 0 | 1 |
| *110* | 1 | 0 | 1 | 1 | 1 | 1 |
| *101* | 1 | 1 | 0 | 1 | 1 | 1 |
| *100* | 1 | 1 | 1 | 0 | 1 | 1 |
| *001* | 0 | 1 | 1 | 1 | 0 | 1 |
| *111* | 1 | 1 | 1 | 1 | 1 | 0 |

In a similar way, for the data exemplars we have the outcome vector **Ω** (defined in section 2.5 and discussed in section 2.7, especially section 2.7.2, of Skousen 2005):

| *y* | *x* | *x* | *x* | *x* | *x* |
|---|---|---|---|---|---|
| 0 | 1 | 1 | 1 | 1 | 1 |

But we now do a pairwise comparison of the individual outcome vector, resulting in the outcome difference array **W**×**W** (or more simply **W**$^2$). This array will be the same for each supracontext:

>        from $j = 1$ to $m$ do
>            from $j' = 1$ to $m$ do
>                IDENTITY ($\Omega^{[j]}$, $\Omega^{[j']}$, $Y^2_{jj'} = 1$, $W^2_{jj'} = 1$)

**$W^2$** (1 means the outcomes are different, 0 the same)

|   | 0 | 1 | 1 | 1 | 1 | 1 |
|---|---|---|---|---|---|---|
| 0 | 0 | 1 | 1 | 1 | 1 | 1 |
| 1 | 1 | 0 | 0 | 0 | 0 | 0 |
| 1 | 1 | 0 | 0 | 0 | 0 | 0 |
| 1 | 1 | 0 | 0 | 0 | 0 | 0 |
| 1 | 1 | 0 | 0 | 0 | 0 | 0 |
| 1 | 1 | 0 | 0 | 0 | 0 | 0 |

The next stage is to perform a reversible conjunction on the two arrays, $V^2$ and $W^2$, producing an array $P \times P$ (or more simply $P^2$) that shows the heterogeneity of the directional pointers between each data exemplar:

>        from $j = 1$ to $m$ do
>            from $j' = 1$ to $m$ do
>                CCNOT($V^2_{jj'}$, $W^2_{jj'}$, $P^2_{jj'} = 0$)

**$P^2$** (1 means that both the subcontext and the outcome are different, 0 otherwise)

| $V^2$ | | | | | | $W^2$ | | | | | | ⇛ | $P^2$ | | | | | |
|---|---|---|---|---|---|---|---|---|---|---|---|---|---|---|---|---|---|---|
| 0 | 1 | 1 | 1 | 0 | 1 | 0 | 1 | 1 | 1 | 1 | 1 | | 0 | 1 | 1 | 1 | 0 | 1 |
| 1 | 0 | 1 | 1 | 1 | 1 | 1 | 0 | 0 | 0 | 0 | 0 | | 1 | 0 | 0 | 0 | 0 | 0 |
| 1 | 1 | 0 | 1 | 1 | 1 | 1 | 0 | 0 | 0 | 0 | 0 | | 1 | 0 | 0 | 0 | 0 | 0 |
| 1 | 1 | 1 | 0 | 1 | 1 | 1 | 0 | 0 | 0 | 0 | 0 | | 1 | 0 | 0 | 0 | 0 | 0 |
| 0 | 1 | 1 | 1 | 0 | 1 | 1 | 0 | 0 | 0 | 0 | 0 | | 0 | 0 | 0 | 0 | 0 | 0 |
| 1 | 1 | 1 | 1 | 1 | 0 | 1 | 0 | 0 | 0 | 0 | 0 | | 1 | 0 | 0 | 0 | 0 | 0 |

Given these preliminary arrays, we now create the containment array for each supracontext (which will be done simultaneously in our quantum superpositioning of the supracontexts). In section 2.6 of the Skousen 2005 paper, I showed how to construct the containment vector $\mathbb{C}$, which is a vector of $m$ qubits, one for each exemplar in the data set. In this paper, I construct a corresponding containment array $\mathbb{C} \times \mathbb{C}$ (or more simply $\mathbb{C}^2$). $\mathbb{C}^2$ is initially set to zero (that is, $\mathbb{C}^2 = \mathbb{0}^2$), and we work through the array changing any qubit in $\mathbb{C}^2$, say $\mathbb{C}^2_{jj'}$, to $\mathbb{1}$ whenever its pair of difference vectors, $\mathbf{D}^{[j]}$ and $\mathbf{D}^{[j']}$, is in the supracontext $\mathbb{S}$:

```
from j = 1 to m do
{   INCLUSION($\mathbb{S}$, $\mathbf{D}^{[j]}$, $\mathbb{X}$ = 1, $\mathbb{Y}$ = 0);
    from j' = 1 to m do
    {   INCLUSION($\mathbb{S}$, $\mathbf{D}^{[j']}$, $\mathbb{W}$ = 1, $\mathbb{Z}$ = 0);
        CCNOT($\mathbb{Y}_n$, $\mathbb{Z}_n$, $\mathbb{C}^2_{jj'}$);
        INCLUSION$^{-1}$($\mathbb{S}$, $\mathbf{D}^{[j']}$, $\mathbb{W}$, $\mathbb{Z}$)   }
    INCLUSION$^{-1}$($\mathbb{S}$, $\mathbf{D}^{[j]}$, $\mathbb{X}$, $\mathbb{Y}$)   }
```

I now apply this procedure to the example described in section 1.2 of Skousen 2005. The resulting supracontextual array $\mathbb{C}^2$ shows for each supracontext the possible directional pointers between pairs of exemplars and specifies whether those pointers are in the supracontext or not:

| 111 | 0 0 0 0 0 0 | 100 | 1 0 0 0 1 0 |
|---|---|---|---|
|  | 0 0 0 0 0 0 |  | 0 0 0 0 0 0 |
|  | 0 0 0 0 0 0 |  | 0 0 0 0 0 0 |
|  | 0 0 0 0 0 0 |  | 0 0 0 0 0 0 |
|  | 0 0 0 0 0 0 |  | 1 0 0 0 1 0 |
|  | 0 0 0 0 0 0 |  | 0 0 0 0 0 0 |
| 110 | 1 0 0 0 1 0 | 010 | 1 0 1 1 1 0 |
|  | 0 0 0 0 0 0 |  | 0 0 0 0 0 0 |
|  | 0 0 0 0 0 0 |  | 1 0 1 1 1 0 |
|  | 0 0 0 0 0 0 |  | 1 0 1 1 1 0 |
|  | 1 0 0 0 1 0 |  | 1 0 1 1 1 0 |
|  | 0 0 0 0 0 0 |  | 0 0 0 0 0 0 |
| 101 | 0 0 0 0 0 0 | 001 | 0 0 0 0 0 0 |
|  | 0 0 0 0 0 0 |  | 0 1 0 1 0 0 |
|  | 0 0 0 0 0 0 |  | 0 0 0 0 0 0 |
|  | 0 0 0 0 0 0 |  | 0 1 0 1 0 0 |
|  | 0 0 0 0 0 0 |  | 0 0 0 0 0 0 |
|  | 0 0 0 0 0 0 |  | 0 0 0 0 0 0 |
| 011 | 0 0 0 0 0 0 | 000 | 1 1 1 1 1 1 |
|  | 0 0 0 0 0 0 |  | 1 1 1 1 1 1 |
|  | 0 0 0 0 0 0 |  | 1 1 1 1 1 1 |
|  | 0 0 0 1 0 0 |  | 1 1 1 1 1 1 |
|  | 0 0 0 0 0 0 |  | 1 1 1 1 1 1 |
|  | 0 0 0 0 0 0 |  | 1 1 1 1 1 1 |

For each supracontext, we perform a reversible conjunction for each pairwise pointer in $\mathbb{C}^2$ against $\mathbf{P}^2$, which gives us the heterogeneity array, $\mathbb{H}\times\mathbb{H}$ (or more simply $\mathbb{H}^2$), for that supracontext:

```
from j = 1 to m do
    from j' = 1 to m do
        CCNOT($\mathbb{C}^2_{jj'}$, $\mathbf{P}^2_{jj'}$, $\mathbb{H}^2_{jj'}$ = 0)
```

$\mathbb{H}^2$ (1 means that the pointer in the supracontext is heterogeneous, 0 homogeneous)

|     | $\mathbb{C}^2$ | $\mathbf{P}^2$ | $\Rightarrow$ | $\mathbb{H}^2$ |
|-----|----------------|----------------|---------------|----------------|
| 111 | 0 0 0 0 0 0<br>0 0 0 0 0 0<br>0 0 0 0 0 0<br>0 0 0 0 0 0<br>0 0 0 0 0 0<br>0 0 0 0 0 0 | 0 1 1 1 0 1<br>1 0 0 0 0 0<br>1 0 0 0 0 0<br>1 0 0 0 0 0<br>0 0 0 0 0 0<br>1 0 0 0 0 0 | | 0 0 0 0 0 0<br>0 0 0 0 0 0<br>0 0 0 0 0 0<br>0 0 0 0 0 0<br>0 0 0 0 0 0<br>0 0 0 0 0 0 |
| 110 | 1 0 0 0 1 0<br>0 0 0 0 0 0<br>0 0 0 0 0 0<br>0 0 0 0 0 0<br>1 0 0 0 1 0<br>0 0 0 0 0 0 | 0 1 1 1 0 1<br>1 0 0 0 0 0<br>1 0 0 0 0 0<br>1 0 0 0 0 0<br>0 0 0 0 0 0<br>1 0 0 0 0 0 | | 0 0 0 0 0 0<br>0 0 0 0 0 0<br>0 0 0 0 0 0<br>0 0 0 0 0 0<br>0 0 0 0 0 0<br>0 0 0 0 0 0 |
| 101 | 0 0 0 0 0 0<br>0 0 0 0 0 0<br>0 0 0 0 0 0<br>0 0 0 0 0 0<br>0 0 0 0 0 0<br>0 0 0 0 0 0 | 0 1 1 1 0 1<br>1 0 0 0 0 0<br>1 0 0 0 0 0<br>1 0 0 0 0 0<br>0 0 0 0 0 0<br>1 0 0 0 0 0 | | 0 0 0 0 0 0<br>0 0 0 0 0 0<br>0 0 0 0 0 0<br>0 0 0 0 0 0<br>0 0 0 0 0 0<br>0 0 0 0 0 0 |
| 011 | 0 0 0 0 0 0<br>0 0 0 0 0 0<br>0 0 0 0 0 0<br>0 0 0 1 0 0<br>0 0 0 0 0 0<br>0 0 0 0 0 0 | 0 1 1 1 0 1<br>1 0 0 0 0 0<br>1 0 0 0 0 0<br>1 0 0 0 0 0<br>0 0 0 0 0 0<br>1 0 0 0 0 0 | | 0 0 0 0 0 0<br>0 0 0 0 0 0<br>0 0 0 0 0 0<br>0 0 0 0 0 0<br>0 0 0 0 0 0<br>0 0 0 0 0 0 |
| 100 | 1 0 0 0 1 0<br>0 0 0 0 0 0<br>0 0 0 0 0 0<br>0 0 0 0 0 0<br>1 0 0 0 1 0<br>0 0 0 0 0 0 | 0 1 1 1 0 1<br>1 0 0 0 0 0<br>1 0 0 0 0 0<br>1 0 0 0 0 0<br>0 0 0 0 0 0<br>1 0 0 0 0 0 | | 0 0 0 0 0 0<br>0 0 0 0 0 0<br>0 0 0 0 0 0<br>0 0 0 0 0 0<br>0 0 0 0 0 0<br>0 0 0 0 0 0 |
| 010 | 1 0 1 1 1 0<br>0 0 0 0 0 0<br>1 0 1 1 1 0<br>1 0 1 1 1 0<br>1 0 1 1 1 0<br>0 0 0 0 0 0 | 0 1 1 1 0 1<br>1 0 0 0 0 0<br>1 0 0 0 0 0<br>1 0 0 0 0 0<br>0 0 0 0 0 0<br>1 0 0 0 0 0 | | 0 0 1 1 0 0<br>0 0 0 0 0 0<br>1 0 0 0 0 0<br>1 0 0 0 0 0<br>0 0 0 0 0 0<br>0 0 0 0 0 0 |

|       | $\mathbb{C}^2$        | $\mathbf{P}^2$        | $\Rightarrow$ | $\mathbb{H}^2$        |
|-------|-----------------------|-----------------------|---------------|-----------------------|
| 001   | 0 0 0 0 0 0           | 0 1 1 1 0 1           |               | 0 0 0 0 0 0           |
|       | 0 1 0 1 0 0           | 1 0 0 0 0 0           |               | 0 0 0 0 0 0           |
|       | 0 0 0 0 0 0           | 1 0 0 0 0 0           |               | 0 0 0 0 0 0           |
|       | 0 1 0 1 0 0           | 1 0 0 0 0 0           |               | 0 0 0 0 0 0           |
|       | 0 0 0 0 0 0           | 0 0 0 0 0 0           |               | 0 0 0 0 0 0           |
|       | 0 0 0 0 0 0           | 1 0 0 0 0 0           |               | 0 0 0 0 0 0           |
|       |                       |                       |               |                       |
| 000   | 1 1 1 1 1 1           | 0 1 1 1 0 1           |               | 0 1 1 1 0 1           |
|       | 1 1 1 1 1 1           | 1 0 0 0 0 0           |               | 1 0 0 0 0 0           |
|       | 1 1 1 1 1 1           | 1 0 0 0 0 0           |               | 1 0 0 0 0 0           |
|       | 1 1 1 1 1 1           | 1 0 0 0 0 0           |               | 1 0 0 0 0 0           |
|       | 1 1 1 1 1 1           | 0 0 0 0 0 0           |               | 0 0 0 0 0 0           |
|       | 1 1 1 1 1 1           | 1 0 0 0 0 0           |               | 1 0 0 0 0 0           |

We now test for the heterogeneity of each supracontext. Basically, we are looking for any ones in the supracontext. This means that we are looking for any heterogeneous pointer in $\mathbb{H}^2$. Equivalently, we could negate the entire $\mathbb{H}^2$ and look for any zeros. If the entire negated $\mathbb{H}^2$ was ones, then $\mathbb{H}^2$ would be homogeneous. We use a revised version of the ONES operator on the negated $\mathbb{H}^2$ to hunt for any zeros. (See section 2.3.2 in Skousen 2005 for the original version of the ONES operator.) We do this as follows:

   NOT($\mathbb{H}^2$);
   ONES($\mathbb{H}^2$, $\mathbb{F}_0^2 = 1$, $\mathbb{F}^2 = 0$)

The revised ONES operator uses an $\mathbb{F}^2$ array, initially set to all zeros, to go through the negated $\mathbb{H}^2$ hunting for any zero. There is a triggering qubit $\mathbb{F}_0^2$ outside the $\mathbb{F}^2$ array that is initially set to one and is used to start the hunt for the first zero in $\mathbb{H}^2$. For convenience, we use an index $k$ to show where we linearly are in the array, so that $(j, j')$ is at $k = (j - 1) m + j'$. Thus $k$ starts out in the array at 1 and goes to $m^2$, where $m$ is the number of data points (that is, exemplars). In other words, the index goes from 1 to $m^2$. As an example, $\mathbb{F}_{3,4}^2$ (in the third row and fourth column) is linearly at position $k = 2m + 4 = 16$ (given that $m$, the number of exemplars, is 6). Using $k$, we thus specify the revised operation ONES($\mathbb{H}^2$, $\mathbb{F}_0^2 = 1$, $\mathbb{F}^2 = 0$) as follows:

   from $k = 1$ to $m^2$ do
      CCNOT($\mathbb{H}_k^2$, $\mathbb{F}_{k-1}^2$, $\mathbb{F}_k^2$)

We first check $\mathbb{F}_0^2$, our initial qubit, against $\mathbb{H}_1^2$. As noted earlier, $\mathbb{F}_0^2$ is set at one. So if $\mathbb{H}_1^2$ is a one, then $\mathbb{F}_1^2$ is changed to a one. But if $\mathbb{H}_1^2$ is a zero, then $\mathbb{F}_1^2$ stays as a zero. We continue in this way, moving through the negated $\mathbb{H}^2$ hunting for zeros. If we find at least one zero, the last qubit in $\mathbb{F}^2$ (the one in position $m^2$) will be set at zero, which means that the negated $\mathbb{H}^2$ had a zero and thus the original $\mathbb{H}^2$ had at least a one and was heterogeneous. On the other hand, if the last qubit in $\mathbb{F}^2$ is set at one, there was no zero in the negated $\mathbb{H}^2$, and thus the original $\mathbb{H}^2$ had only zeros and was therefore homogeneous.

I now show in our example how this specifically applies to each supracontext of $\mathbb{H}^2$. Alongside the contain array $\mathbb{C}^2$, I show the negated $\mathbb{H}^2$ and the final state of $\mathbb{F}^2$, with the last qubit in $\mathbb{F}^2$ specified as $\mathbb{F}_{m,m}^2$ (this last qubit in $\mathbb{F}^2$, in the bottom right-hand corner, will be underlined).

A zero for $\mathbb{F}^2_{m,m}$ means that the original $\mathbb{H}^2$ was homogeneous; a one means that $\mathbb{H}^2$ was heterogeneous:

|  | $\mathbb{C}^2$ | NOT($\mathbb{H}^2$) | $\mathbb{F}^2$ |
|---|---|---|---|
| 111 | 0 0 0 0 0 0<br>0 0 0 0 0 0<br>0 0 0 0 0 0<br>0 0 0 0 0 0<br>0 0 0 0 0 0<br>0 0 0 0 0 0 | 1 1 1 1 1 1<br>1 1 1 1 1 1<br>1 1 1 1 1 1<br>1 1 1 1 1 1<br>1 1 1 1 1 1<br>1 1 1 1 1 1 | 0 0 0 0 0 0<br>0 0 0 0 0 0<br>0 0 0 0 0 0<br>0 0 0 0 0 0<br>0 0 0 0 0 0<br>0 0 0 0 0 <u>0</u> |
| 110 | 1 0 0 0 1 0<br>0 0 0 0 0 0<br>0 0 0 0 0 0<br>0 0 0 0 0 0<br>1 0 0 0 1 0<br>0 0 0 0 0 0 | 1 1 1 1 1 1<br>1 1 1 1 1 1<br>1 1 1 1 1 1<br>1 1 1 1 1 1<br>1 1 1 1 1 1<br>1 1 1 1 1 1 | 0 0 0 0 0 0<br>0 0 0 0 0 0<br>0 0 0 0 0 0<br>0 0 0 0 0 0<br>0 0 0 0 0 0<br>0 0 0 0 0 <u>0</u> |
| 101 | 0 0 0 0 0 0<br>0 0 0 0 0 0<br>0 0 0 0 0 0<br>0 0 0 0 0 0<br>0 0 0 0 0 0<br>0 0 0 0 0 0 | 1 1 1 1 1 1<br>1 1 1 1 1 1<br>1 1 1 1 1 1<br>1 1 1 1 1 1<br>1 1 1 1 1 1<br>1 1 1 1 1 1 | 0 0 0 0 0 0<br>0 0 0 0 0 0<br>0 0 0 0 0 0<br>0 0 0 0 0 0<br>0 0 0 0 0 0<br>0 0 0 0 0 <u>0</u> |
| 011 | 0 0 0 0 0 0<br>0 0 0 0 0 0<br>0 0 0 0 0 0<br>0 0 0 1 0 0<br>0 0 0 0 0 0<br>0 0 0 0 0 0 | 1 1 1 1 1 1<br>1 1 1 1 1 1<br>1 1 1 1 1 1<br>1 1 1 1 1 1<br>1 1 1 1 1 1<br>1 1 1 1 1 1 | 0 0 0 0 0 0<br>0 0 0 0 0 0<br>0 0 0 0 0 0<br>0 0 0 0 0 0<br>0 0 0 0 0 0<br>0 0 0 0 0 <u>0</u> |
| 100 | 1 0 0 0 1 0<br>0 0 0 0 0 0<br>0 0 0 0 0 0<br>0 0 0 0 0 0<br>1 0 0 0 1 0<br>0 0 0 0 0 0 | 1 1 1 1 1 1<br>1 1 1 1 1 1<br>1 1 1 1 1 1<br>1 1 1 1 1 1<br>1 1 1 1 1 1<br>1 1 1 1 1 1 | 0 0 0 0 0 0<br>0 0 0 0 0 0<br>0 0 0 0 0 0<br>0 0 0 0 0 0<br>0 0 0 0 0 0<br>0 0 0 0 0 <u>0</u> |

|  | $\mathbb{C}^2$ | NOT($\mathbb{H}^2$) | $\mathbb{F}^2$ |
|---|---|---|---|
| 010 | 1 0 1 1 1 0 | 1 1 0 0 1 1 | 0 0 1 1 1 1 |
|  | 0 0 0 0 0 0 | 1 1 1 1 1 1 | 1 1 1 1 1 1 |
|  | 1 0 1 1 1 0 | 0 1 1 1 1 1 | 1 1 1 1 1 1 |
|  | 1 0 1 1 1 0 | 0 1 1 1 1 1 | 1 1 1 1 1 1 |
|  | 1 0 1 1 1 0 | 1 1 1 1 1 1 | 1 1 1 1 1 1 |
|  | 0 0 0 0 0 0 | 1 1 1 1 1 1 | 1 1 1 1 1 <u>1</u> |
| 001 | 0 0 0 0 0 0 | 1 1 1 1 1 1 | 0 0 0 0 0 0 |
|  | 0 1 0 1 0 0 | 1 1 1 1 1 1 | 0 0 0 0 0 0 |
|  | 0 0 0 0 0 0 | 1 1 1 1 1 1 | 0 0 0 0 0 0 |
|  | 0 1 0 1 0 0 | 1 1 1 1 1 1 | 0 0 0 0 0 0 |
|  | 0 0 0 0 0 0 | 1 1 1 1 1 1 | 0 0 0 0 0 0 |
|  | 0 0 0 0 0 0 | 1 1 1 1 1 1 | 0 0 0 0 0 <u>0</u> |
| 000 | 1 1 1 1 1 1 | 1 0 0 0 1 0 | 0 1 1 1 1 1 |
|  | 1 1 1 1 1 1 | 0 1 1 1 1 1 | 1 1 1 1 1 1 |
|  | 1 1 1 1 1 1 | 0 1 1 1 1 1 | 1 1 1 1 1 1 |
|  | 1 1 1 1 1 1 | 0 1 1 1 1 1 | 1 1 1 1 1 1 |
|  | 1 1 1 1 1 1 | 1 1 1 1 1 1 | 1 1 1 1 1 1 |
|  | 1 1 1 1 1 1 | 0 1 1 1 1 1 | 1 1 1 1 1 <u>1</u> |

Note that for each supracontext we must go through the entire negated array, NOT($\mathbb{H}^2$), in order to determine its heterogeneity. We cannot stop the process as soon as we find the first zero since some supracontexts will be homogeneous and will thus require us to make a search through the entire negated array to make sure there is no zero. The quantum superposition of the supracontexts requires the same operators to be applied to each supracontext from beginning to end.

Before we reverse the ONES operation and restore the original array $\mathbb{H}^2$, we need to apply the value in the last qubit of $\mathbb{F}^2$ (namely, $\mathbb{F}^2_{m,m}$), thus eliminating the pointers in heterogeneous supracontexts. For each supracontext, we create the analogy array, $\mathbb{A}^2$, which is initially set at all zeros (that is, $\mathbb{A}^2 = \mathbb{0}^2$). If $\mathbb{F}^2_{m,m}$ is a zero, then the supracontext is homogeneous and $\mathbb{A}^2$ is made identical to $\mathbb{C}^2$. On the other hand, if $\mathbb{F}^2_{m,m}$ is a one, then the supracontext is heterogeneous and $\mathbb{A}^2$ remains as $\mathbb{0}^2$. For each supracontext, we negate $\mathbb{F}^2_{m,m}$ and perform the operation CCNOT($\mathbb{C}^2$, $\mathbb{F}^2_{m,m}$, $\mathbb{A}^2 = \mathbb{0}^2$) for each ($j$, $j'$):

NOT( $\mathbb{F}^2_{m,m}$ );
from $j = 1$ to $m$ do
    from $j' = 1$ to $m$ do
        CCNOT($\mathbb{C}^2_{jj'}$, $\mathbb{F}^2_{m,m}$, $\mathbb{A}^2_{jj'} = \mathbb{0}$)

So for each ($j$, $j'$) we therefore have two cases each for homogeneity and heterogeneity:

| | $\mathbb{F}^2_{m,m}$ | $\mathbb{C}^2_{jj'}$ | NOT($\mathbb{F}^2_{m,m}$) | $\mathbb{A}^2_{jj'} = \mathbb{0}$ | CCNOT($\mathbb{C}^2$, $\mathbb{F}^2_{m,m}$, $\mathbb{A}^2_{jj'}$) |
|---|---|---|---|---|---|
| *homogeneity* | 0 | 0 | 1 | 0 | 0 |
| | 0 | 1 | 1 | 0 | 1 |
| *heterogeneity* | 1 | 0 | 0 | 0 | 0 |
| | 1 | 1 | 0 | 0 | 0 |

Thus we can specifically create the analogical array for each supracontext:

| | $\mathbb{F}^2_{m,m}$ | $\mathbb{C}^2$ | ⇛ | $\mathbb{A}^2$ | *homogeneous pointers* |
|---|---|---|---|---|---|
| 111 | 0 | 0 0 0 0 0 0 | | 0 0 0 0 0 0 | |
| | | 0 0 0 0 0 0 | | 0 0 0 0 0 0 | |
| | | 0 0 0 0 0 0 | | 0 0 0 0 0 0 | |
| | | 0 0 0 0 0 0 | | 0 0 0 0 0 0 | |
| | | 0 0 0 0 0 0 | | 0 0 0 0 0 0 | |
| | | 0 0 0 0 0 0 | | 0 0 0 0 0 0 | |
| 110 | 0 | 1 0 0 0 1 0 | | 1 0 0 0 1 0 | oms / y → oms / y |
| | | 0 0 0 0 0 0 | | 0 0 0 0 0 0 | oms / y → omn / x |
| | | 0 0 0 0 0 0 | | 0 0 0 0 0 0 | |
| | | 0 0 0 0 0 0 | | 0 0 0 0 0 0 | |
| | | 1 0 0 0 1 0 | | 1 0 0 0 1 0 | omn / x → oms / y |
| | | 0 0 0 0 0 0 | | 0 0 0 0 0 0 | omn / x → omn / x |
| 101 | 0 | 0 0 0 0 0 0 | | 0 0 0 0 0 0 | |
| | | 0 0 0 0 0 0 | | 0 0 0 0 0 0 | |
| | | 0 0 0 0 0 0 | | 0 0 0 0 0 0 | |
| | | 0 0 0 0 0 0 | | 0 0 0 0 0 0 | |
| | | 0 0 0 0 0 0 | | 0 0 0 0 0 0 | |
| | | 0 0 0 0 0 0 | | 0 0 0 0 0 0 | |
| 011 | 0 | 0 0 0 0 0 0 | | 0 0 0 0 0 0 | |
| | | 0 0 0 0 0 0 | | 0 0 0 0 0 0 | |
| | | 0 0 0 0 0 0 | | 0 0 0 0 0 0 | |
| | | 0 0 0 1 0 0 | | 0 0 0 1 0 0 | cma / x → cma / x |
| | | 0 0 0 0 0 0 | | 0 0 0 0 0 0 | |
| | | 0 0 0 0 0 0 | | 0 0 0 0 0 0 | |

|  | $\mathbb{F}^2_{m,m}$ | $\mathbb{C}^2$ | | | | | $\Rightarrow$ | $\mathbb{A}^2$ | | | | | | *homogeneous pointers* |
|---|---|---|---|---|---|---|---|---|---|---|---|---|---|---|
| 100 | 0 | 1 | 0 | 0 | 0 | 1 | 0 | | 1 | 0 | 0 | 0 | 1 | 0 | oms / y → oms / y |
|  |  | 0 | 0 | 0 | 0 | 0 | 0 | | 0 | 0 | 0 | 0 | 0 | 0 | oms / y → omn / x |
|  |  | 0 | 0 | 0 | 0 | 0 | 0 | | 0 | 0 | 0 | 0 | 0 | 0 |  |
|  |  | 0 | 0 | 0 | 0 | 0 | 0 | | 0 | 0 | 0 | 0 | 0 | 0 |  |
|  |  | 1 | 0 | 0 | 0 | 1 | 0 | | 1 | 0 | 0 | 0 | 1 | 0 | omn / x → oms / y |
|  |  | 0 | 0 | 0 | 0 | 0 | 0 | | 0 | 0 | 0 | 0 | 0 | 0 | omn / x → omn / x |
| 010 | 1 | 1 | 0 | 1 | 1 | 1 | 0 | | 0 | 0 | 0 | 0 | 0 | 0 |  |
|  |  | 0 | 0 | 0 | 0 | 0 | 0 | | 0 | 0 | 0 | 0 | 0 | 0 |  |
|  |  | 1 | 0 | 1 | 1 | 1 | 0 | | 0 | 0 | 0 | 0 | 0 | 0 |  |
|  |  | 1 | 0 | 1 | 1 | 1 | 0 | | 0 | 0 | 0 | 0 | 0 | 0 |  |
|  |  | 1 | 0 | 1 | 1 | 1 | 0 | | 0 | 0 | 0 | 0 | 0 | 0 |  |
|  |  | 0 | 0 | 0 | 0 | 0 | 0 | | 0 | 0 | 0 | 0 | 0 | 0 |  |
| 001 | 0 | 0 | 0 | 0 | 0 | 0 | 0 | | 0 | 0 | 0 | 0 | 0 | 0 |  |
|  |  | 0 | 1 | 0 | 1 | 0 | 0 | | 0 | 1 | 0 | 1 | 0 | 0 | gfa / x → gfa / x |
|  |  | 0 | 0 | 0 | 0 | 0 | 0 | | 0 | 0 | 0 | 0 | 0 | 0 | gfa / x → cma / x |
|  |  | 0 | 1 | 0 | 1 | 0 | 0 | | 0 | 1 | 0 | 1 | 0 | 0 | cma / x → gfa / x |
|  |  | 0 | 0 | 0 | 0 | 0 | 0 | | 0 | 0 | 0 | 0 | 0 | 0 | cma / x → cma / x |
|  |  | 0 | 0 | 0 | 0 | 0 | 0 | | 0 | 0 | 0 | 0 | 0 | 0 |  |
| 000 | 1 | 1 | 1 | 1 | 1 | 1 | 1 | | 0 | 0 | 0 | 0 | 0 | 0 |  |
|  |  | 1 | 1 | 1 | 1 | 1 | 1 | | 0 | 0 | 0 | 0 | 0 | 0 |  |
|  |  | 1 | 1 | 1 | 1 | 1 | 1 | | 0 | 0 | 0 | 0 | 0 | 0 |  |
|  |  | 1 | 1 | 1 | 1 | 1 | 1 | | 0 | 0 | 0 | 0 | 0 | 0 |  |
|  |  | 1 | 1 | 1 | 1 | 1 | 1 | | 0 | 0 | 0 | 0 | 0 | 0 |  |
|  |  | 1 | 1 | 1 | 1 | 1 | 1 | | 0 | 0 | 0 | 0 | 0 | 0 |  |

The homogenous pointers are all listed above in the right-hand column. Randomly selecting one of the these, we get to choose from 4 pointers to the *y* outcome and 9 pointers to the *x* outcome. Thus the probability of choosing the *y* outcome is 4/13.

As explained in section 2.8 of Skousen 2005, we can observe a particular outcome in QAM by randomly selecting any one of the pointers in any of the homogeneous supracontexts. Each homogeneous pointer is equally possible, no matter which homogeneous supracontext it is found in. In other words, we do not have to first collapse the superposition and then chose a homogeneous pointer in the supracontext that has been selected (the traditional approach in quantum mechanics).

When we compare this revised procedure with the one originally given in section 2.7 of Skousen 2005, we can see how much simpler it is to determine the homogeneity of a supracontext by hunting for heterogeneous pointers than by determining whether all the exemplars contained in a supracontext show a plurality of subcontexts and a plurality of outcomes. But more importantly, we are using directional pointers to minimize the uncertainty of the resulting prediction. These pointers directly represent the disagreement, the quadratic measure of uncertainty.

*Acknowledgments*

I wish to thank my colleague Deryle Lonsdale for his careful reading of this paper and for his helpful comments and suggestions.